\documentclass[%
 reprint,
superscriptaddress,
 amsmath,amssymb,
 aps,
floatfix,
]{revtex4-2}

\usepackage{graphicx}
\usepackage{float}
\usepackage{units}
\usepackage{amsmath}
\usepackage{amssymb}
\usepackage{braket}
\usepackage{bm}
\usepackage{gensymb}
\usepackage[utf8]{inputenc}
\usepackage{xcolor}
\usepackage[T1]{fontenc}
\newcommand\vect[1]{\mathbf{#1}}
\newcommand\ten[1]{\hat{#1}}

\usepackage{booktabs}
\usepackage{tabularx}

\usepackage{array}

\usepackage{hyperref}

\begin{document}

\preprint{APS/123-QED}

\title{In-situ control of hole-spin driving mechanisms}

\author{Simon~Geyer}
\thanks{These authors contributed equally to this work}
\affiliation{Department of Physics, University of Basel, Klingelbergstrasse 82, \\ CH-4056 Basel, Switzerland}

\author{Rafael~S.~Eggli}
\thanks{These authors contributed equally to this work}
\affiliation{Department of Physics, University of Basel, Klingelbergstrasse 82, \\ CH-4056 Basel, Switzerland}

\author{Carlos~Dos~Santos}
\affiliation{Department of Physics, University of Basel, Klingelbergstrasse 82, \\ CH-4056 Basel, Switzerland}

\author{Miguel~J.~Carballido}
\thanks{Present address: School of Electrical Engineering and Telecommunications, University of New South Wales, Sydney, NSW 2052, Australia}
\affiliation{Department of Physics, University of Basel, Klingelbergstrasse 82, \\ CH-4056 Basel, Switzerland}

\author{Peter~Stano}
\affiliation{RIKEN, Center for Quantum Computing (RQC), 2-1 Hirosawa, Wako, Saitama 351-0198, Japan}
\affiliation{Institute of Physics, Slovak Academy of Sciences, 845 11 Bratislava, Slovakia}

\author{Daniel~Loss}
\thanks{Present address: Physics Department, King Fahd University of Petroleum and Minerals (KFUPM), 31261, Dhahran, Saudi Arabia. Center for Advanced Quantum Computing, KFUPM, Dhahran, Saudi Arabia. RDIA Chair in Quantum Computing}
\affiliation{Department of Physics, University of Basel, Klingelbergstrasse 82, \\ CH-4056 Basel, Switzerland}
\affiliation{RIKEN, Center for Quantum Computing (RQC), 2-1 Hirosawa, Wako, Saitama 351-0198, Japan}

\author{Dominik~M.~Zumb\"uhl}
\affiliation{Department of Physics, University of Basel, Klingelbergstrasse 82, \\ CH-4056 Basel, Switzerland}

\author{Richard~J.~Warburton}
\affiliation{Department of Physics, University of Basel, Klingelbergstrasse 82, \\ CH-4056 Basel, Switzerland}

\author{Andreas~V.~Kuhlmann}
\email{e-mail: andreas.kuhlmann@unibas.ch}
\affiliation{Department of Physics, University of Basel, Klingelbergstrasse 82, \\ CH-4056 Basel, Switzerland}

\date{\today}

\begin{abstract}
Hole-spin qubits enable fast, all-electrical spin manipulation through electric-dipole spin resonance (EDSR), arising from two microscopic mechanisms rooted in their intrinsically strong spin-orbit interaction. Depending on how the electric field acts on the quantum dot, the spin can be driven either by a modulation of its $g$-factor or by a displacement of the wavefunction. Here, we demonstrate in-situ control over the dominant EDSR driving mechanism of a hole-spin qubit in a silicon fin field-effect transistor by applying microwave signals to two different gate electrodes, thereby tuning the orientation of the local electric field. We measure the effective $g$-factor, its electrical tunability, and the Rabi frequency as functions of magnetic-field orientation. Their distinct angular dependencies, analyzed using a $g$-matrix formalism, allow us to identify the underlying driving processes and track their relative contributions for different drive configurations. By selecting the drive electrode, we can switch from a regime dominated by $g$-factor modulation to one with a strong contribution from wavefunction displacement. This in-situ tunability provides direct experimental access to both spin-driving mechanisms and offers a route toward optimized spin-qubit performance.
\end{abstract}

\maketitle

\section{Introduction}
Spin qubits in semiconductor quantum dots (QDs) are promising candidates for high-fidelity, scalable quantum computing \cite{Burkard2023,Loss1998,Bulaev2005,Philips2022,GonzalezZalba2021,Weinstein2023,Neyens2024,George2024,Steinacker2024,Wu2025,Huang2024}. In comparison to electron spins, hole spins \cite{Maurand2016,Hendrickx2020a,Hendrickx2020,Hendrickx2021,Camenzind2022,Piot2022,Hendrickx2024,Geyer2024,Bassi2024,Vorreiter2025,Froning2021a} can be controlled all-electrically via EDSR \cite{Golovach2006,Bulaev2007,Kloeffel2018,Bosco2021a}, without the added complexity of on-chip micromagnets~\cite{Tokura2006,PioroLadriere2008} or the need for an orbital degeneracy~\cite{Gilbert2023}, thanks to their intrinsic spin-orbit interaction (SOI). In materials with strong SOI, such as holes in silicon (Si), the Larmor (spin precession) vector $\vect{f}_\mathrm{L}$ and magnetic field $\vect{B}$ are no longer related via a scalar $g$-factor. Instead, a 3x3 $g$-tensor $\ten{g}$ \cite{Sen2023,Sharma2024,Liles2021,Geyer2024,Hendrickx2024} is required, leading to $\vect{f}_\mathrm{L} = \mu_\mathrm{B}\ten{g}\vect{B}/h$, where $\mu_B$ is Bohr's magneton and $h$ is Planck's constant. To induce spin rotations using EDSR, a periodic modulation of $\vect{f}_\mathrm{L}$ transverse to the spin quantization axis $\vect{n}=\ten{g}\vect{B}/|\ten{g}\vect{B}|$ must be applied close to the qubit's Larmor frequency $f_\mathrm{L} = |\vect{f}_\mathrm{L}|$.

This can be achieved through two fundamentally different mechanisms \cite{Crippa2018,Venitucci2018,Michal2021}. The first one is referred to as $g$-tensor modulation resonance (gTMR), in which a transverse modulation of the Larmor vector arises because an alternating (ac) electric field deforms the $g$-tensor -- either by rotating its principal axes or by changing its principal values in an anisotropic manner~\cite{Ares2013} (Fig.~\ref{fig1}f). Electrical control of $\ten{g}$ is enabled by SOI, primarily through changes in the heavy- and light-hole mixing~\cite{Ares2013a} induced by the confinement potential~\cite{Pryor2006}, as well as by strain or strain gradients~\cite{Liles2021}.

Secondly, coherent spin rotations can be driven by periodically displacing the QD as a whole with an ac electric field~\cite{Golovach2006,Nowack2007} (Fig.~\ref{fig1}g). In the presence of SOI, this motion generates an effective time-dependent magnetic field, resulting in a non-collinear modulation of the Larmor vector. For a quasi-harmonic confinement, the hole wave function and the effective $g$-factor $g^* = |\ten{g}\vect{B}|/|\vect{B}|$, and hence the Zeeman splitting $E_\mathrm{Z} = h f_L$, remain unchanged during the motion. Therefore, this driving mechanism is called iso-Zeeman EDSR (IZR)~\cite{Crippa2018}.


In practice, the driving mechanisms typically coexist, and different relative weights of gTMR and IZR have been reported for various hole-spin qubit devices~\cite{Crippa2018,Hendrickx2024,Carballido2025}. To describe these mechanisms within a unified framework, the $g$-matrix formalism was developed~\cite{Venitucci2018,Michal2021,Crippa2018}, which enables the individual contributions to the observed Rabi drive to be disentangled.

The importance of the underlying driving mechanism is highlighted by an operational sweet spot with simultaneously maximized qubit drive speed and coherence~\cite{Carballido2025} as a function of the static electric field, made possible by predominantly IZR-based driving. Likewise, gTMR-based driving can produce ``reciprocal sweetness'' \cite{Bassi2024,Mauro2024,Michal2023}, where qubit speed and coherence peak together as a function of the magnetic-field direction. While these results open promising avenues for qubit optimizations, in-situ electrical control over the relative contributions of the two mechanisms to the overall Rabi frequency has not been achieved to date. Such control ultimately depends on how the microwave electric field couples to the QD~\cite{Michal2021}, which in turn is determined by the layout of the gate electrodes and the geometry of the confinement potential. For example, driving from a more remote gate electrode~\cite{Carballido2025} can promote a lateral displacement of the QD along the transport direction -- the axis of weak confinement -- in systems with strong biaxial confinement~\cite{Camenzind2022,Maurand2016,Nowack2007}, thereby enhancing the IZR contribution. Experimental confirmation and quantification of how the relative position of the driving gate electrode with respect to the qubit affects the driving mechanism are therefore needed.

Here, we investigate the driving mechanisms of a hole-spin qubit in a silicon fin field-effect transistor (FinFET) \cite{Geyer2021,Geyer2024,Camenzind2022,Eggli2025,Bosco2023}, modeling the experimental results using the $g$-matrix formalism. By applying the microwave drive signal to two distinct gate electrodes, we achieve in-situ control over the EDSR driving mechanism, tuning the system from a gTMR-dominated regime to one in which IZR contributes substantially to the Rabi frequency. These results demonstrate that the position of the driving gate relative to the qubit plays a decisive role in determining the dominant driving mechanism, opening opportunities for device designs that enables real-time, on-demand switching between gTMR and IZR operation. Finally, we discuss the physical origin of this tunability and its connection to sweet-spot operation.

\section{Experimental Setup}
\begin{figure*}[htb!]
    \centering
    \includegraphics[width=120mm]{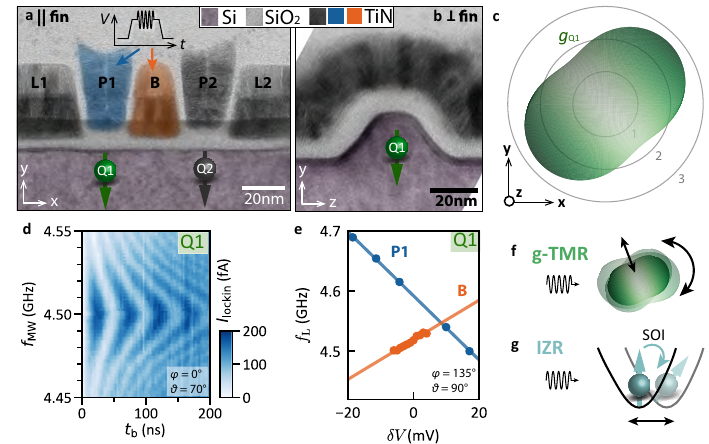}
    \caption{\textbf{Hole spin qubit anisotropies and driving mechanisms.} \textbf{a}, False-color transmission electron microscope (TEM) image of the measured device, showing the cross-section along the fin direction. The qubits (Q1, Q2) are located beneath the plunger gates (P1, P2) and are manipulated by applying microwave signals to either P1 or the barrier gate B. The barrier voltage controls the inter-dot tunneling, and the lead gates (L1, L2) accumulate the hole reservoirs. The widths of the B and P gates are $\unit[{\simeq}\,{20}]{nm}$. \textbf{b}, Cross-sectional TEM image of a co-fabricated device, highlighting the nearly triangular fin shape. \textbf{c}, 3D representation of the $g$-tensor, showing a large anisotropy within the $xy$-plane. \textbf{d}, Rabi chevron pattern of Q1 under P1 drive, demonstrating a Rabi frequency of $f_\mathrm{R}=20$\,MHz on resonance. Data were taken at $B=0.136$\,T, $\phi = 0^\circ$, $\theta = 70^\circ$ (as defined in Fig.~\ref{fig2}), and $V_\mathrm{MW,P1}=12$\,mV. \textbf{e}, The qubit Larmor frequency $f_\mathrm{L}$ of Q1 is tuned by the voltages applied to P1 and B. Data were taken at $B=0.172$\,T, $\phi = 135^\circ$, $\theta = 90^\circ$. \textbf{f, g}, Schematic illustrations of the gTMR and IZR driving mechanisms, respectively.}
    \label{fig1}
\end{figure*}

Two hole QDs are formed in the Si fin beneath plunger gates P1 and P2, each hosting an effective spin-1/2 qubit labeled Q1 and Q2 (see Fig.~\ref{fig1}a). The fin has a nearly triangular cross-section (see Fig.~\ref{fig1}b), which produces quasi-1D confinement of the holes and induces a strong direct-Rashba SOI~\cite{Kloeffel2018, Bosco2021a}. 
A square pulse applied to P1 switches the device between Pauli spin blockade (PSB) \cite{Ono2002}, used for spin initialization and spin-parity readout via direct-current transport measurements, and Coulomb blockade~\cite{Nowack2007}. In the latter regime, fast microwave bursts are applied for spin manipulation via EDSR. These  bursts can be applied to either P1 or the barrier gate B. A difference in the effective $g$-factors allows the two qubits to be addressed independently. To enhance spin readout contrast we employ a lock-in detection scheme~\cite{Camenzind2022,Maurand2016,Geyer2024,Froning2021a}, yielding the signal $I_\mathrm{lockin}$. A three-axis vector magnet enables experiments with an arbitrarily oriented magnetic field.

\section{Results}
\subsection{Anisotropy of the g-tensor}
We characterize $\ten{g}$ \cite{Sen2023,Sharma2024} of Q1 by probing its Zeeman splitting spectroscopically. As the orientation of $\vect{B}$ is varied, the anisotropy of $\ten{g}$ \cite{Bosco2021,Winkler2003,Jin2024,John2024,Liles2021} appears through $E_\mathrm{Z} = \mu_\mathrm{B}|\ten{g}\vect{B}|$ (Supplemental Material Fig.~S1) \cite{Geyer2024}. Using the procedure of Crippa et al.~\cite{Crippa2018} (Supplemental Material Section~A), we obtain

\begin{eqnarray}
    \ten{g} = \left(
\begin{array}{ccc}
 ~~2.38 & ~~0.52 & -0.01 \\
 ~~0.52 & ~~2.22 & ~~0.01 \\
 -0.01 & ~~0.01 & ~~1.58 \\
\end{array}
\right),
\end{eqnarray}

as visualized in Fig.~\ref{fig1}c. Although $\ten{g}$ contains nine elements, spectroscopy provides access only to the magnitude of the effective Zeeman field $|\ten{g}\vect{B}|$, not its direction. Consequently, only six independent parameters can be determined: three principal $g$-factors, which set the Zeeman splitting along the principal magnetic axes, and three Euler angles specifying their orientation~\cite{Crippa2018,Geyer2024,Liles2021}. The remaining three degrees of freedom correspond to an arbitrary rotation in spin space -- equivalently, a unitary transformation of the Kramers basis -- which leaves $|\ten{g}\vect{B}|$ and thus all observables invariant~\cite{Crippa2018,Mauro2024}. 

In the following, we present the results for Q1; a corresponding analysis for Q2 is given in Supplemental Material Sections~B and F. We also compare the extracted $g$-tensor with a microscopic model (Supplemental Material Section~C), which offers a physical interpretation of the observed anisotropy.

\subsection{Anisotropy of the Rabi frequency}
\begin{figure*}[t]
    \centering
    \includegraphics[width=\textwidth]{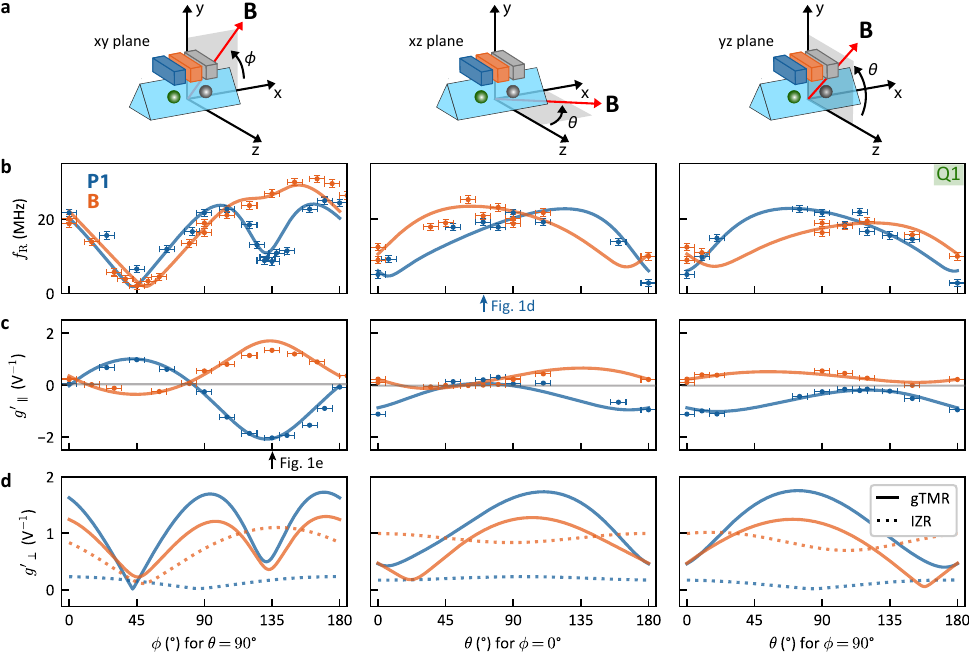}
    \caption{\textbf{Rabi drive anisotropy.} \textbf{a}, Schematic of the three magnetic-field planes used in the experiments, shown relative to the orientation of the Si fin (light blue) and the two driving gate electrodes (dark blue, orange). \textbf{b}, Rabi frequency of Q1 under P1- (blue markers) and B-driving (orange markers) as a function of magnetic-field orientation. Data were taken at fixed $f_\mathrm{L}=4.5\,$GHz, $V_\mathrm{MW,P1}=12\,$mV, and $V_\mathrm{MW,B}=13.5\,$mV. Solid curves are fits based on the model described in the main text. The uncertainty in $f_\mathrm{R}$ is estimated as $\pm1$\,MHz. \textbf{c}, Electrical longitudinal tunability of the Q1 $g$-factor as function of magnetic-field orientation for P1- (blue markers) and B-driving (orange markers). Solid curves correspond to model fits described in the main text. \textbf{d}, Solid and dashed curves show the transverse components of $\ten{g}'$, as defined in the main text. The 1$\sigma$ error bars in $g'$ were obtained by propagating an estimated magnetic field hysteresis of $\pm5$\,mT. Gaps in the data are due to failing PSB readout for some magnetic field orientations.} 
    \label{fig2}
\end{figure*}

Next, we drive Q1 for a variable burst duration $t_\mathrm{b}$ from P1 using a microwave amplitude $V_\mathrm{MW,P1}=12$\,mV and a frequency $f_\mathrm{MW}$ close to $f_\mathrm{L}=4.5$\,GHz (Fig.~\ref{fig1}d). The Rabi frequency $f_\mathrm{R}$ is extracted by fitting $I_\mathrm{lockin}$ at resonance as a function of $t_\mathrm{b}$, and this procedure is repeated for different magnetic-field orientations in the three planes shown in Fig.~\ref{fig2}a. The resulting data (Fig.~\ref{fig2}b, blue markers) reveal a pronounced angular dependence of $f_\mathrm{R}$, with certain orientations exhibiting strongly suppressed Rabi driving (e.g. $\phi~\simeq~45\degree; \theta~=~90\degree$). Driving the qubit from the barrier gate B with $V_\mathrm{MW,B}=13.5$\,mV instead leads to a markedly different anisotropy (Fig.~\ref{fig2}b, orange markers). To understand this behavior, we now examine the underlying EDSR driving mechanisms.

If the driving electric field distorts the QD's confinement potential, it can deform the $g$-tensor, giving rise to gTMR-driven spin rotations~\cite{Kato2003,Ares2013}. The magnitude of the $g$-factor tunability, defined as $g'=\partial g/\partial V$, is determined by measuring the linear response of $f_\mathrm{L}$ to small gate-voltage changes $\delta V$ (Fig.\ \ref{fig1}e). Together with the measured $\ten{g}$ and the drive amplitude $V_\mathrm{MW}$, the matrix describing this tunability -- denoted as the ``gTMR-induced component of the $g$-derivative,''  $\ten{g}'_\mathrm{gTMR}$ -- fully specifies the gTMR-induced Rabi frequency $f_\mathrm{R,gTMR}$~\cite{Carballido2025} for any magnetic-field orientation~\cite{Crippa2018}.

In contrast, IZR originates from a displacement of the QD in a system with SOI~\cite{Golovach2006,Nowack2007}, while the confinement potential and therefore $g^*$ and $E_\mathrm{Z}$ remain unchanged. Within the $g$-matrix formalism ~\cite{Crippa2018,Venitucci2018,Gyoergy2025}, IZR accounts for the difference between the experimentally observed $f_\mathrm{R}$ and the value predicted from gTMR alone, yielding $f_\mathrm{R,IZR}$. Analogous to gTMR, it is represented by the ``IZR-induced component of the $\ten{g}$ derivative,'' $\ten{g}'_\mathrm{IZR}$.

Rabi driving arising from a mixture of both mechanisms is described by~\cite{Crippa2018}:
\begin{equation}
    \vect{f}_\mathrm{R} = \frac{\mu_\mathrm{B}V_\mathrm{MW}|\vect{B}|}{2h} \left[ \vect{n} \times \left( \ten{g}'\vect{b} \right)\right], \quad  \vect{b} =\frac{\vect{B}}{|\vect{B}|}.
    \label{eq:Rabi_main}
\end{equation}
Here, $\vect{f}_\mathrm{R}$ defines the Rabi vector, whose magnitude gives the Rabi frequency $f_\mathrm{R}=|\vect{f}_\mathrm{R}|$, while its direction corresponds to the axis of spin rotation on the Bloch sphere. The quantity $\ten{g}' = \ten{g}'_\mathrm{gTMR} + \ten{g}'_\mathrm{IZR}$ denotes the derivative of the $g$-tensor with respect to the gate voltage on the driving electrode, with contributions from gTMR and IZR. The term $\ten{g}'\vect{b}$ represents the oscillating effective magnetic field, and the cross product with the spin-quantization axis $\vect{n}$ ensures that only components perpendicular to the static Zeeman field contribute to the Rabi drive. Note that both driving mechanisms scale linearly with the electric- and magnetic-field amplitudes, and therefore cannot be separated without analyzing their distinct angular dependencies (Supplemental Section~D).

\subsection{Extraction of g-tunability and full Rabi-model reconstruction}

Next, we determine $\ten{g}'_\mathrm{gTMR}$ and $\ten{g}'_\mathrm{IZR}$ following the method of Crippa \emph{et al.}~\cite{Crippa2018} (Supplemental Section~E). We begin by measuring how $E_\mathrm{Z}$ changes with the voltage applied to the driving gate (Fig.~\ref{fig2}c), yielding the longitudinal $g$-factor tunability $\frac{\partial E_\mathrm{Z}}{\partial V} = \mu_\mathrm{B}Bg'_\mathrm{\parallel}$, where $g'_\mathrm{\parallel} = \frac{\partial |\ten{g}\vect{b}|}{\partial V} = \left[\ten{g}'\vect{b}\right]\cdot\vect{n}$. Here, $g'_\mathrm{\parallel}$ quantifies the voltage-induced change in the magnitude of the effective Zeeman field, i.e., its component along the spin-quantization axis. Consequently, $g'_\mathrm{\parallel}$ modulates the Zeeman splitting without tilting $\vect{n}$, producing dephasing rather than coherent rotations. Because IZR leaves $E_\mathrm{Z}$ invariant, it contributes no longitudinal response, so $\ten{g}'_\mathrm{IZR,\parallel} = 0$. Thus, the measured $g'_\mathrm{\parallel}$ reflects only the gTMR contribution, enabling a reconstruction of $\ten{g}'_\mathrm{gTMR}$ -- up to the usual unitary freedom in the choice of spin basis -- for driving from gates P1 and B~\cite{Crippa2018}. Projecting $\ten{g}'_\mathrm{gTMR}$ onto the spin-quantization axis and magnetic-field orientation according to $g'_\mathrm{\parallel} = \left[\ten{g}'_\mathrm{gTMR}\vect{b}\right]\cdot \vect{n}$ reproduces the measured data with good agreement (Fig.\ \ref{fig2} c, solid curves), thereby validating the reconstruction.

The transverse component of this tensor governs coherent Rabi driving by gTMR and is given by $g'_\mathrm{gTMR,\perp}=\lvert\left[\ten{g}'_\mathrm{gTMR}\vect{b}\right]\times\vect{n}\rvert$. By comparing the predicted $f_\mathrm{R,gTMR}$ with the measured $f_\mathrm{R}$, the remaining contribution can be attributed to IZR, which allows extraction of $\ten{g}'_\mathrm{IZR}$~\cite{Crippa2018}. By definition, $\ten{g}'_\mathrm{IZR}$ has no longitudinal component but a finite transverse response $g'_\mathrm{IZR,\perp} = \lvert\left[\ten{g}'_\mathrm{IZR}\vect{b}\right]\times\vect{n}\rvert$, leading to coherent spin rotations. 

Utilizing Eq.~\eqref{eq:Rabi_main} together with the determined $\ten{g}' = \ten{g}'_\mathrm{gTMR} + \ten{g}'_\mathrm{IZR}$, we obtain a model that incorporates both driving mechanisms and is fully consistent with the measured angular dependence of the Rabi frequency (Fig.~\ref{fig2}b, solid curves). The matrices for driving from P1 are
\begin{eqnarray}
\ten{g}'{}_\mathrm{gTMR}^\mathrm{P1} = \left(
\begin{array}{ccc}
 -0.36 & ~~1.52 & ~~0.31 \\
 ~~1.55 & -0.71 & -0.35 \\
 ~~0.35 & -0.37 & -0.87 \\
\end{array}
\right) \mathrm{V}^{-1},
\end{eqnarray}

\begin{eqnarray}
\ten{g}'{}_\mathrm{IZR}^\mathrm{P1}=\left(
\begin{array}{ccc}
 ~~0.006 & ~~0.019 & -0.163 \\
 -0.022 & -0.004 & ~~0.049 \\
 -0.300 & -0.015 & -0.002 \\
\end{array}
\right) \mathrm{V}^{-1},
\end{eqnarray}
and for driving from B
\begin{eqnarray}
\ten{g}'{}_\mathrm{gTMR}^\mathrm{B} = \left(
\begin{array}{ccc}
 ~~0.62 & -1.01 & -0.40 \\
 -1.03 & ~~0.71 & ~~0.27 \\
 -0.47 & ~~0.16 & ~~0.20 \\
\end{array}
\right) \mathrm{V}^{-1},
\end{eqnarray}

\begin{eqnarray}
\ten{g}'{}_\mathrm{IZR}^\mathrm{B}=\left(
\begin{array}{ccc}
 -0.015 & ~~0.118 & ~~0.725 \\
 -0.071 & -0.038 & -0.681 \\
 -0.832 & ~~0.717 & ~~0.053 
\end{array}
\right) \mathrm{V}^{-1}.
\end{eqnarray}

\begin{figure*}[t]
    \centering
    \includegraphics[width=\textwidth]{fig3.pdf}
    \caption{\textbf{IZR and gTMR Rabi drive contributions.} \textbf{a}, Calculated Rabi frequency of Q1 (first column), decomposed into the contributions from IZR (second column) and gTMR (third column) under P1-driving. The dominant contribution arises from gTMR, with $p_\mathrm{gTMR}=85\%$. \textbf{b}, Same analysis for B-driving. Here, IZR and gTMR contribute comparably to the total Rabi driving ($p_\mathrm{IZR}=55\%$ and $p_\mathrm{gTMR}=45\%$). Constructive (destructive) interference between the two mechanisms occurs at the red $+$ ($-$) symbols, leading to enhanced (suppressed) overall Rabi frequency.}
    \label{fig3}
\end{figure*}

We now compare the $g$-factor tunability when driving from gates P1 and B. Across the two drive configurations, $g'_\mathrm{\parallel}$ shows an angular dependence that is mirrored about zero (Fig.~\ref{fig2}c, solid curves). In contrast, $g'_\mathrm{gTMR,\perp}$ follows a similar angular trend for both drive configurations (Fig.~\ref{fig2}d, solid curves). Notably, the magnitudes of $g'_\mathrm{\parallel}$ and $g'_\mathrm{gTMR,\perp}$ are generally smaller under B-driving (Fig.~\ref{fig2}c and d, orange curves) than under P1-driving (blue curves), which is counterintuitive given the larger gate lever arm of gate B $\alpha^\mathrm{B} = 0.208\pm0.005$ versus $\alpha^\mathrm{P1} = 0.125\pm0.002$, indicating stronger electric-field coupling. Conversely, B-driving results in a much larger $g'_\mathrm{IZR,\perp}$ (Fig.~\ref{fig2}d, orange dotted curves) compared to P1-driving (blue dotted curves). This finding highlights the nontrivial nature of $g$-tuning by gate voltages, as it pertains to effects beyond simple capacitive couplings.

\subsection{Tuning the EDSR drive composition}
Our model enables a quantitative comparison of the driving contributions from the two mechanisms under P1- or B-driving by computing the Rabi vector $\vect{f}_\mathrm{R}$ and its components $\vect{f}_\mathrm{R,IZR}$ and $\vect{f}_\mathrm{R,gTMR}$ for all magnetic-field orientations (see Fig.~\ref{fig3}). The latter two ``pure'' contributions are obtained directly from Eq.~\eqref{eq:Rabi_main} by substituting $\ten{g}'$ with the corresponding matrix for each driving mechanism. Since the total Rabi vector is given by $\vect{f}_\mathrm{R} = \vect{f}_\mathrm{R,IZR} + \vect{f}_\mathrm{R, gTMR}$, their magnitudes satisfy the triangle inequality $f_\mathrm{R} \le f_\mathrm{R,IZR} + f_\mathrm{R,gTMR}$. The two contributions add vectorially and, depending on their relative orientation, may interfere constructively, destructively, or anywhere in between. From the calculated maps of $f_\mathrm{R, gTMR}$ and $f_\mathrm{R, IZR}$, it is evident that gTMR dominates the Rabi driving under P1-driving, whereas B-driving produces a markedly different pattern, with both mechanisms contributing significantly and exhibiting distinct anisotropies.

To quantify this observation, we define the average fractional contribution of each mechanism as:
\begin{align}
p_\mathrm{IZR} = \frac{\langle|\vect{f}_\mathrm{R,IZR}|\rangle}{\mathcal{F}}, \quad
p_\mathrm{gTMR} =  \frac{\langle|\vect{f}_\mathrm{R,gTMR}|\rangle}{\mathcal{F}}, \\ \mathcal{F} = \langle|\vect{f}_\mathrm{R,IZR}|\rangle+\langle|\vect{f}_\mathrm{R,gTMR}|\rangle,
\end{align}
where $\langle|\vect{x}|\rangle$ denotes the average magnitude of $\vect{x}$ over all magnetic-field orientations, i.e., over the solid angle of the unit sphere. The results are presented in Table \ref{tab:avg_drive_strength_Q1}, confirming the conclusions drawn above.

\begin{table}[htbp]
\centering
\caption{Drive contributions for Q1.}
\begin{tabular}{c @{\hskip 2cm} c @{\hskip 2cm} c}
\toprule
 & $p^\mathrm{Q1}_\mathrm{IZR}$ & $p^\mathrm{Q1}_\mathrm{gTMR}$ \\
\midrule
P1 drive & 15\% & 85\% \\
B drive & 55\% & 45\% \\
\bottomrule
\end{tabular}
\label{tab:avg_drive_strength_Q1}
\end{table}

 Thus, selecting different gate electrodes enables \textit{in-situ} control over the relative contributions of the two EDSR driving mechanisms. Notably, the maximum $f_\mathrm{R,IZR}$ increases from below $4$\,MHz under P1-driving to above $20$\,MHz under B-driving -- a fivefold enhancement achieved by laterally shifting the drive source by only $\simeq\,30$\,nm. This demonstrates a high degree of tunability and highlights the potential for further optimization of the device geometry to tailor the spin-driving mechanism.

\section{Discussion}
\subsection{Opportunities arising from drive tuning}
While the distinct anisotropies of gTMR and IZR allow the relative contributions of the two mechanisms to be tuned by adjusting the magnetic-field orientation~\cite{Crippa2018}, switching the driving gate offers an all-electrical means of control that can be executed much faster than rotating the external field. Gate switching also enables local tuning at the level of individual qubits, rather than relying on the global magnetic-field orientation. In principle, such electrical control could be automated and carried out adaptively using fast feedback~\cite{Huang2024}, enabling real-time adjustment of the driving mechanism. Furthermore, by applying the same microwave tone simultaneously to both the B and P1 gates \cite{John2024,Gyoergy2025} and controlling the relative amplitudes and phases on the two gates, one could smoothly transition from a gTMR-dominated regime to configurations where gTMR and IZR contribute comparably. 

Additionally, the static gate voltages can be adjusted, shifting the QD position in the Si channel~\cite{Bassi2024} and tuning its exchange interaction~\cite{Geyer2024} while also affecting $\ten{g}$ and $\ten{g}'$~\cite{Liles2021,Froning2021} and thus the qubit drive composition~\cite{Carballido2025}, presenting yet another lever to pull in future experiments.

\subsection{Link between electric-field orientation and driving mechanism}
The relative strength of the two EDSR driving mechanisms is governed by the orientation of the microwave electric field~\cite{Michal2021}. An ac electric field applied along a weak confinement direction predominantly displaces the QD without significantly altering the confinement potential, thereby activating IZR~\cite{Golovach2006,Carballido2025}. In contrast, a field aligned with a strong confinement axis induces minimal displacement but deforms the orbital wavefunction, giving rise to gTMR~\cite{Ares2013}. 

The behavior observed in our device is consistent with this interpretation. Q1 lies beneath gate P1; thus, driving from P1 primarily generates out-of-plane electric-field lines, which mainly deform the confinement potential and therefore favor gTMR. By contrast, gate B is laterally offset from the qubit, producing at Q1 an ac electric field that contains not only a vertical component but also a substantial in-plane component along the fin - the direction of weak confinement. This naturally results in a mixture of IZR and gTMR. Accordingly, we observe an enhanced IZR contribution when the driving gate is laterally displaced from the position directly above the qubit, consistent with an increased in-plane electric-field component. This trend aligns with recent reports of predominantly IZR driving using more distant gates in nanowire devices~\cite{Carballido2025}.

\subsection{Optimizing qubit operation}
From a qubit-performance perspective, $f_\mathrm{R}$ is a key metric as it determines the achievable qubit gate speed. In our device, the maximum individual contributions from both gTMR and IZR reach $\simeq\,25$\,MHz (see Fig.~\ref{fig3}). This symmetry likely reflects the specific device geometry; in general, the maximum Rabi frequency achievable via each mechanism can be engineered. For example, IZR can be enhanced by elongating the QD along the fin, since the IZR-induced Rabi frequency is predicted to scale with the fourth power of the confinement length along the direction of motion~\cite{Froning2021}, enabling ultra-fast qubit gate operations~\cite{Bosco2021b,Michal2021}.

An interesting observation, consistent with previous reports~\cite{Crippa2018,Bassi2024}, is the existence of multiple magnetic-field orientations for which $f_\mathrm{R}\,\simeq\,0$ under gTMR driving, whereas only a single such orientation exists for IZR - namely when $\vect{B}$ lies parallel to the spin-orbit field direction $\vect{n_\mathrm{SO}}$~\cite{Golovach2006,Nowack2007,Liles2024}. In the context of scaling to larger qubit arrays~\cite{Philips2022,George2024,Seidler2025}, achieving uniform qubit gate speeds across all qubits is essential. Although gTMR-based high-speed field orientations can be electrically tuned on a per-qubit basis~\cite{Bassi2024}, IZR may offer a practical advantage due to its narrower low-speed regime, making it easier to avoid orientations with vanishing Rabi frequency.

Another important metric is the dephasing time $T_2^*$. Since dephasing occurs during qubit idling - when no drive is applied - it is independent of the chosen drive configuration. In general, achieving optimal Rabi speeds while simultaneously maximizing coherence is not guaranteed, as reducing sensitivity to noise typically also suppresses coupling to the driving field~\cite{Piot2022}. For predominantly gTMR-driven hole spin qubits, the concept of ``reciprocal sweetness'' has been proposed to mitigate this trade-off~\cite{Michal2023,Mauro2024,Bassi2024}. It predicts that local maxima of $f_\mathrm{R}$ as a function of magnetic-field orientation should coincide with coherence sweet spots. Our FinFET qubit exhibits indications of this relation in the P1-driving configuration, where gTMR dominates (see Supplemental Section G). Notably, for $\vect{B}$ aligned with $[\phi,\theta] = [10^\circ, 120^\circ]$, both $T_2^*$ and $f_\mathrm{R}$ reach a local maximum. In contrast, driving from gate B shifts the magnetic-field orientation associated with the highest Rabi frequency to a point that no longer coincides with a coherence sweet spot. This suggests that ``reciprocal sweetness'' is favored in regimes where gTMR is the dominant driving mechanism.

In light of this result, IZR may appear to be a less favorable driving mechanism when aiming to optimize both gate speed and coherence. However, simultaneous maximization of these two metrics can also be achieved in predominantly IZR-driven systems that employ an elongated quantum dot~\cite{Carballido2025}. In such systems, the hole $g$-factor is renormalized and suppressed by increasing spin-orbit coupling strength~\cite{Froning2021}, and IZR causes $f_\mathrm{R}$ to monotonically rise with the spin-orbit coupling~\cite{Bosco2021b,Kloeffel2018}. Importantly, the strength of the direct-Rashba spin-orbit interaction peaks at a finite electric field resulting in a simultaneously maximal $f_\mathrm{R}$ and vanishing derivative of $g$ with respect to gate voltage \cite{Carballido2025}, thus additionally ensuring a coherence sweet spot~\cite{Bosco2021a,Kloeffel2018}. Although our experiments do not reach an IZR-dominated regime, the FinFET architecture naturally favors elongated quantum dots, and tuning the gate voltages may shift the drive composition further toward IZR. It also remains an open question whether an analogous sweet spot condition for both speed and coherence exists in regimes with a more balanced drive composition.

Ultimately, these considerations suggest that ``pure'' driving using either IZR or gTMR facilitates the realization of fast and coherent hole spin qubits. We demonstrate that \textit{in-situ} switching from a less desirable mixed configuration to a much more ``pure'' drive composition is achievable. Because qubits located at different sites in a larger quantum dot array may reach their optimal drive under varying conditions, the search and consecutive adaptive adjustment of the drive configuration promises to be a useful tool for optimizing future hole spin processors.

\section{Conclusion}
In conclusion, we have demonstrated \textit{in-situ} switching between two distinct driving configurations of a hole spin qubit in a Si FinFET. By mapping the $g$-tensor, its derivatives with respect to plunger- and barrier-gate voltages, and the Rabi frequency, and fitting all quantities using the $g$-matrix formalism, we obtain a quantitative description of the underlying driving composition. When the drive is applied to the plunger gate P1, the qubit is predominantly driven via gTMR, resulting from an ac electric modulation of the $g$-tensor. In contrast, driving from the barrier gate B yields a more balanced drive composition, with significant contributions from both gTMR and IZR, the latter originating from lateral QD displacement induced by the ac electric field. Switching from P1- to B-driving enhances the IZR-induced Rabi contribution by a factor of five. These findings are consistent with the driving field orientation being governed by the spatial position of the driving gate relative to the QD. Importantly, switching the driving mechanism by selecting the driving gate is an all-electrical process, eliminating the need for slow tuning of the magnetic-field orientation.

\section{Outlook}
Having demonstrated in-situ switching between Rabi driving mechanisms, the individual influence of gTMR and IZR on qubit operation can now be systematically explored. In the broader context of operating spin qubits in a quantum processor, where numerous processes beyond Rabi driving occur (e.g., idling, shuttling~\cite{Bosco2024,Struck2024}, resonator coupling~\cite{Mi2018,Samkharadze2018,Landig2018,Dijkema2024}, two-qubit gates~\cite{Huang2024,Hendrickx2020,Geyer2024} and readout~\cite{Oakes2023,Kiyama2024}), differences between gTMR and IZR may offer distinct advantages or pose specific challenges.

Looking ahead, the ability to electrically switch the driving mechanism may serve as a powerful diagnostic tool for spin qubits and could be used to evaluate the impact of design modifications in future device architectures. Owing to the strong dependence of the IZR contribution on the direction of the ac field, such devices could also be exploited to locally probe the microwave field orientation or, conversely, to triangulate the position of the QD relative to a radiation source.

While our results clearly demonstrate that a single device can be tuned from a gTMR-dominated configuration toward a more balanced regime between gTMR and IZR, they also raise important questions for future work. In particular, it remains to be investigated whether a more complete inversion of the drive composition - i.e.\ a transition from $p_\mathrm{gTMR} \gg p_\mathrm{IZR}$ to $p_\mathrm{gTMR} \ll p_\mathrm{IZR}$ - can be achieved. This may be possible by applying the drive to the right plunger gate (P2), or to a gate electrode positioned even more laterally along the fin. Furthermore, monochromatic driving from multiple gates with tunable relative amplitudes and phases~\cite{Gyoergy2025} could enable smooth traversal between different driving regimes, providing a platform to study this transition in a controlled manner.

\section*{Acknowledgment}
We thank T.\ Berger, L.\ Camenzind, S.\ Bosco, and T.\ Patlatiuk for useful discussions and experimental support. We acknowledge the support of the cleanroom operation team, particularly U.\ Drechsler, A.\ Olziersky, and D.\ D.\ Pineda, at the IBM Binnig and Rohrer Nanotechnology Center, as well as technical assistance at the University of Basel by S.\ Martin and M.\ Steinacher. This work was supported by NCCR SPIN, a National Center of Competence in Research funded by the Swiss National Science Foundation (grant number 225153). A.V.K.\ and P.S.\ acknowledge funding within the QuantERA II program, which received support from the EU’s Horizon 2020 research and innovation program under Grant Agreement No.\ 101017733.

\section*{Authors Contributions}
S.G., C.D.S., R.S.E., and A.V.K.\ performed the experiments and analyzed the data with input from R.J.W. and D.M.Z. S.G., R.S.E., and A.V.K.\ wrote the manuscript with contributions from all authors. M.J.C., P.S., and D.L.\ provided valuable input on the conceptual understanding and theoretical analysis. S.G. and A.V.K.\ conceived the project. A.V.K.\ managed the project with support from R.J.W. and D.M.Z.

\end{document}